\documentclass[a4paper]{article}

\usepackage[english]{babel}
\usepackage[utf8x]{inputenc}
\usepackage[T1]{fontenc}

\usepackage[a4paper,top=3cm,bottom=1.25in,left=3cm,right=3cm,marginparwidth=1.75cm]{geometry}

\title{Targeted Estimation of L2 Distance}

\usepackage{amsmath}
\usepackage{float}
\usepackage{amsfonts}
\usepackage{graphicx}
\usepackage[colorinlistoftodos]{todonotes}
\usepackage[colorlinks=true, allcolors=blue]{hyperref}

\begin{document}

\begin{center}
\end{center}
\vspace{1 cm}
\begin{center} \LARGE
Targeted Estimation of L2 Distance Between Densities and its Application to Geo-spatial Data
\end{center}
\vspace{0.5 cm}


\begin{center} \large
\textsc{George Shan$^1$ \hspace{1.3 cm} Mark J. van der Laan$^{1}$}\\
\end{center}
\begin{center}
\emph{$^1$Division of Biostatistics, University of California at Berkeley\\
}
\end{center}
\vspace{4cm}

\begin{center} \large
Abstract
\end{center}

\large
We examine the integrated squared difference, also known as the L2 distance (L2D), between two probability densities. Such a distance metric allows for comparison of differences between pairs of distributions or changes in a distribution over time. We propose a targeted maximum likelihood estimator for this parameter based on samples of independent and identically distributed observations from both underlying distributions. We compare our method to kernel density estimation and demonstrate superior performance for our method with regards to confidence interval coverage rate and mean squared error.

\newpage

\section{Introduction}
\large

Quantification of the difference between two probability densities can be useful. Some methods like the Kolmogorov-Smirnov score and supremum norm give a measure of maximum difference. Others like the Kullback-Leibler divergence (Kullback and Leibler, 1951) measure the divergence of one density from another. For a symmetric measure of overall differences, the L2 distance (L2D), otherwise known as the integrated square difference, can be useful. Here we study estimation of the L2 distance (L2D) between two unknown distributions using independent iid observations from both distributions. We also apply this estimation to make useful inferences from comparison of pairwise density differences in real data. Previous work with application to L2D has been done using different approaches, with applications including unsupervised change point detection and class balance estimation Sugiyama et al (2013). Estimation of the integral square functional has been examined previously, Bickel and Ritov (1988), Gine and Nickl (2008).

L2 distance between two functions is the integral of the squared difference between the two functions, in our case probability density functions;

\[\Psi(P^0,P^1)=\int_x (p^1(x)-p^0(x))^2 dx \]

\noindent
It can provide a useful indication of the size of differences between two distributions. In the first part of this paper we provide a method of targeted estimation of the true L2 distance between two unknown distributions based on samples from those distributions. This estimator is called a targeted maximum likelihood estimator.

Differences between pairs of densities can be compared across many pairs by estimating the L2D for each pair and comparing those estimates. In the second part of this paper we illustrate the use of L2D estimation on a real world data set to identify the most significant changes resulting from an intervention. 

The most straightforward way to approach the problem of L2D estimation is to estimate the densities of our two distributions separately and then to evaluate the L2D by plugging in our two estimates. This leads to the issue of how to deal with smoothing. In general we will operate under a non-parametric model for density estimation. Non-parametric density estimation requires us to smooth our empirical distribution to find a suitable estimate. However, the degree of smoothing depends on some measure of fidelity to the data generating distribution, such as the cross validated log-likelihood loss, or the mean integrated squared error. Finding an appropriate balance between bias and variance when smoothing a non-parametric density estimate does not guarantee the same optimality with regards to the L2D. The result is that a 'good' estimator for the density of a distribution as a whole is not necessarily a 'good' estimator for the L2 distance.

One clear motivation for a better approach is that L2D depends on the difference between two densities, whereas the typical bias/variance trade off for each individual density estimate is only concerned with that particular distribution. Smoothing is inevitable, but the smoothing should be tailored to our parameter of interest. This is the motivation behind targeted estimation. In order to 'target' the L2D, we should consider how this parameter is sensitive to various parts of the distribution.  

\pagebreak

\section{Targeted Estimation of L2 Distance}

Consider random variable $(X,A)$, of which we have $n$ i.i.d. observations from distribution $P_0$ with density $p_0(x,a)$. $X$ is continuous multivariate with density everywhere defined and $A$ is Bernoulli. Denote $P^0$ as $P|(A=0)$ with density $p^0(x)$ and likewise $P^1$ as $P|(A=1)$ with density $p^1(x)$, both conditional distributions of $X$ for any hypothetical distribution $P$ of $X$ and $A$. Our operating model is non-parametric for both conditional densities $p^0(x)$ and $p^1(x)$, respectively, and holds $0<p(A=1)<1$ as arbitrarily fixed. Our estimand is:

\[\Psi(P_0)=\int_x \{p_0(x|A=1)-p_0(x|A=0)\}^2 dx \]

Our approach is to develop a targeted maximum likelihood estimator (TMLE) based on the framework described by van der Laan and Rubin (2006) and in Targeted Learning (van der Laan and Rose, 2011). We will first define the TMLE and then establish that it is an asymptotically efficient estimator.

A TMLE is a two step substitution estimator. The first step is to estimate the true density $p_0$. We call this the initial estimation step, which gives us an initial density estimate $p_n$, with corresponding probability measure $P_n$. In the second step we update $P_n$ to target it towards our parameter. Consider a submodel that includes $P_n$ and which is indexed by a single parameter with the property that the score at $P_n$ is $D^*(P_n)$, where $D^*(P)$ is the canonical gradient of $\Psi$ at $P$. We say that this property is the \textit{least favorable model} (LFM) property and the mentioned submodel possesses this property at $P_n$ so it is a \textit{local least favorable model} (LLFM). There is a unique such submodel through $P_n$ such that the LFM property applies at every member element, not just at $P_n$. We call this the \textit{universal least favorable model} (ULFM). The member of ULFM$(P_n)$ which maximizes the likelihood of our observed data is the TMLE update, $P_n^*$. Finally, the TMLE of $\Psi(P_0)$ is $\Psi(P_n^*)$. van der Laan and Gruber (2016) provide a detailed discussion on these points.

In Appendix A we find the canonical gradient of $\Psi$ to be:
\begin{equation}
\begin{split}
D^*(P)(X,A) &= \frac{I(A=1)}{p(A=1)}* 2*\{p^1(X)-p^0(X) -\int (p^1(x)-p^0(x))dP^1(x)\} \\
           & \quad + \frac{I(A=0)}{p(A=0)}* 2*\{p^0(X)-p^1(X) -\int (p^0(x)-p^1(x))dP^0(x)\} 
\end{split}
\end{equation}

It follows that the second order remainder $R^2(P,P_0)$ defined as $\Psi(P)-\Psi(P_0)+P_0D^*(P)$ = $- \int [(p^1_0-p^0_0)-(p^1-p^0)]^2dx$ \\

We can express the error of our estimator as:
\begin{equation}
\begin{split}
\Psi(P_n^*)-\Psi(P_0) &= (\bold{P}_n-P_0)D^*(P_0) - \bold{P}_nD^*(P_n^*)\\
& \quad +(\bold{P}_n-P_0)\{D^*(P_n^*)-D^*(P_0)\} + R^2(P_n^*,P_0)
\end{split}
\end{equation}

An efficient estimator is a linear estimator with influence function equal to the canonical gradient of the estimand. Its error behaves as $(\bold{P}_n-P_0)D^*(P_0)$, where $\bold{P}_n$ denotes the empirical measure. Such an estimator is the estimator with minimal asymptotic variance among all regularly asymptotically linear estimators.  $\Psi(P_n^*)$ is asymptotically efficient if the last 3 expressions on the right side of Equation 2 converge to zero at a rate faster than $n^{-1/2}$, which leaves the first term, the efficient error, that behaves as a sample mean converging at $n^{-1/2}$ rate. This estimator has Gaussian distribution with mean equal to the estimand and standard error equal to standard deviation of $D^*(P_0)$ divided by $n^{1/2}$. 

If $P_n^*$ is the MLE for a sub-model with score $D^*(P_n^*)$ at $P_n^*$ then the second term in Equation 2 is 0. If $D^*(P_n^*)$ falls in a Donsker class and $P_0(D^*(P_n^*)-D^*(P_0))^2 = o_P(1)$ then the third term on the right side of Equation 2 is $o_P(n^{-1/2})$ as a result of asymptotic equicontinuity of an empirical process indexed by a Donsker class (van der Vaart and Wellner, 1996). Finally, the second order remainder $R^2(P_n^*,P_0)$ is $o_P(n^{-1/2})$ if the density estimator $p_n^*$ converges in L2 norm faster than $n^{-1/4}$ rate. These conditions would imply that $\Psi(P_n^*)-\Psi(P_0) = \bold{P}_nD^*(P_0) + o_P(n^{-1/2})$. This means that under these generally achievable conditions, TMLE is asymptotically efficient. It is asymptotically linear with influence function equal to the canonical gradient of the parameter under the true distribution. It's distribution is asymptotically normal and unbiased. Asymptotically accurate Wald confidence intervals can be constructed centered on $\Psi(P_n^*)$ and using standard deviation of $D^*(P_n^*)$ divided by $n^{1/2}$ as standard error.  Further elaboration on conditions and general satisfiability thereof for asymptotic efficiency of TMLE is provided in Appendix B.\\

For the initial density estimation we will use a Gaussian kernel density estimator with global bandwidth selected by HPI plug in method (Duong, 2007; Wand and Jones, 1994), for the sake of simplicity. In principle any type of regular, consistent estimator can be used, and the most accurate estimator or ensemble for the particular data should be used. It's worth restating that the $R^2$ term depends on the accuracy of the updated density and thus also the initial density estimator. Asymptotic efficiency depends on $R^2$ converging faster than $n^{-1/2}$ and finite sample performance depends on this term as well. 

The update step is the main feature of the TMLE. We find a density estimate that solves its empirical efficient influence equation. There are many such solutions, but we want to be as close to the true distribution as possible. Our best guess at this is our initial estimate. We conceive of a single parameter sub-model that is part of the non-parametric model space and contains our initial estimate as one of its elements. The score of the sub-model at the initial estimate is equal to the canonical gradient applied to that estimate. We call this a `least favorable model', because any perturbation within this parametric model leads to the largest possible change in our parameter (and least favorable to estimation of the estimand). It is `local' in that the canonical gradient applied to the initial estimate is not the same as that applied to a nearby member of the model space, and thus the described property applies to the locality of the initial estimate, but not necessarily to anywhere else. There is one sub-model such that the score at every member is equal to the canonical gradient applied to that member distribution. This we call the `universal least favorable path'. A maximum likelihood estimate along this `universal path' solves its empirical score equation and thus its empirical efficient influence equation. This is the targeted maximum likelihood estimate.

In practice, solving an MLE along the universal path requires repeated small updates starting with the initial estimate. Oftentimes using the gradient at the initial estimate as the score for the entire sub-model will find an MLE in one step that comes close to solving it's empirical efficient influence equation. This is one particular local least favorable model. A rule of thumb for close: if empirical mean of efficient influence function is smaller in magnitude than its empirical standard deviation divided by $n^{1/2}*log(n)$, where $n$ is the sample size. Such a condition would imply that $\bold{P}_nD^*(P_n^*)$ is small enough to not provide a meaningful contribution to the finite sample behavior of the TMLE, while preserving asymptotic efficiency. If this condition is not met, an additional update step can close the gap. In the work discussed here, one update was found to be sufficient to meet this stopping criterion.\\  

The one step TMLE described above was implemented for our simulations and analysis. First, we apply our canonical gradient operator $D^*$ to the initial distribution estimate $P_n$ and obtain $D^*(P_n)$, the score of our local least favorable model.\\

This local least favorable model is $p_{n,\epsilon}$, $\epsilon \in (-\delta,\delta)$:
\[p_{n,\epsilon}(X,A) = (1+\epsilon D^*(P_n)(X,A)) * p_n(X,A)\]

Care should be taken to ensure that perturbation by $\epsilon$ yields a valid density function. This is done by keeping $\delta$ small enough such that the resulting density estimate is not anywhere negative. We now compute the MLE of $\epsilon$ such that it maximizes the log-likelihood:  

\[\epsilon_{MLE} = argmax_\epsilon \sum _ i \log [(1+\epsilon D^*(P_n)(X_i,A_i)) * p_n(X_i,A_i)] \]

Our TMLE update is:

\[p_n^*(X,A) =  (1+\epsilon_{MLE} D^*(P_n)(X,A)) * p_n(X,A)\]

Note that here we use the empirical proportion for $p_n(A)$ and this remains unchanged after the update step. We check that $\bold{P}_nD^*(P_n^*)$ meets our aforementioned criterion of being less than the empirical standard deviation of canonical gradient divided by $n^{1/2}*log(n)$. If so, we plug in $P_n^*$ to $\Psi$ and we have $\Psi(P_n^*)$ as our TMLE of the L2D.

\pagebreak

\section{Simulations: TMLE using initial kernel density estimator}

We study the effect of TMLE update applied to an initial kernel density estimate through simulation. Since the TMLE depends on the initial estimate, and the kernel estimator depends on the regularity of the underlying distribution, we will start off by using three different types of data generating distribution with varying levels of regularity. For each type we will use two overlapping identical distributions with a certain amount of offset corresponding to a true L2 distance. These are Normal distribution with standard deviation 0.5 and 0.5 offset (referred to as 'Gaussian'), Isosceles triangle with base 2 and height 1 offset by 0.5 ('Triangle'), and uniform with range 1 offset by 0.1 ('uniform'). For each type of distribution, we will acquire 5000 simulations of sample size $n$ from both distributions, where $n$ is 50, 100, 200 ... 51200. For each simulation, we will estimate the L2D. This will be done for both kernel density estimates and TMLE adjusted kernel estimates. 

To compare performance, we will assess plug in estimation using both TMLE updated density estimate as well as initial density estimate without update. In both cases we will use Gaussian kernel density estimate with HPI bandwidth. L2D between the kernel density estimates for each distribution is computed using numerical integration. Shown below are comparisons of the estimator performance using both methods, across the range of sample sizes.

For each distribution type we examine confidence interval coverage rate and mean squared error across the range of sample sizes. For coverage rate we compute 95\% intervals that use sample L2D estimate +/- $1.96* SE$. The standard error is calculated using either the standard deviation of the sampling distribution (5000 samples), referred to as 'oracle' since it assumes knowledge that one would not ordinarily have for a single sample, or by using the variance of the empirical influence curve for each sample separately (referred to as 'sample'). The first is the actual standard error, while the second is an estimate. In addition, we examine the mean squared error and variance for the estimates in our sampling distribution, and multiply these figures by the sample size $n$. The variance of the true efficient influence function is shown as a horizontal black line for reference. This is the efficient rate, and any asymptotically efficient estimator should have mean squared error and variance converge to this rate as $n$ increases.\\

\begin{figure}[!ht]
\includegraphics[width=1.9 in]{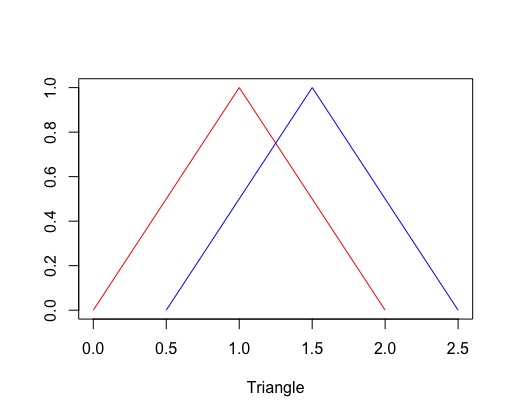}
\includegraphics[width=1.9 in]{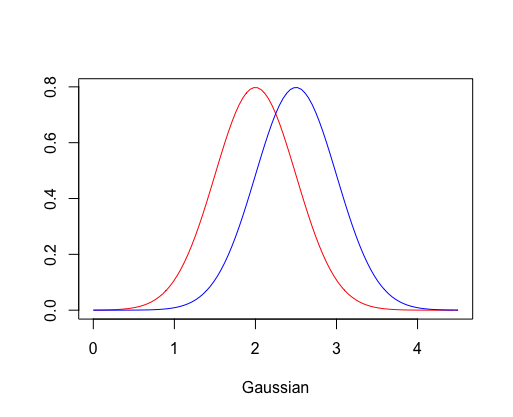}
\includegraphics[width=1.9 in]{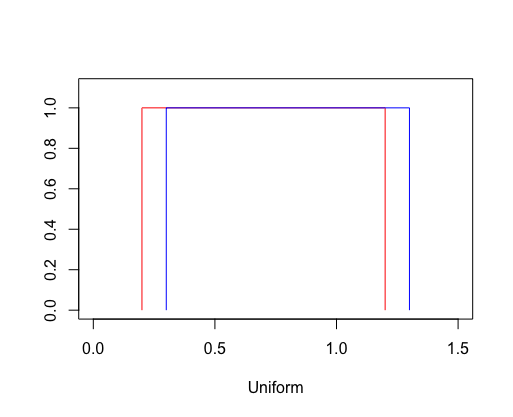}

\caption{Graphs of the density curves of simulated distributions. Red and blue represent the two densities $p^0$ and $p^1$.} 

\end{figure}

\pagebreak

\begin{figure}[!ht]
\includegraphics[width=3 in]{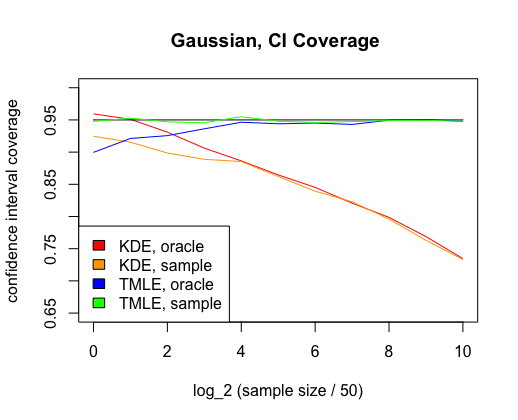}
\includegraphics[width=3 in]{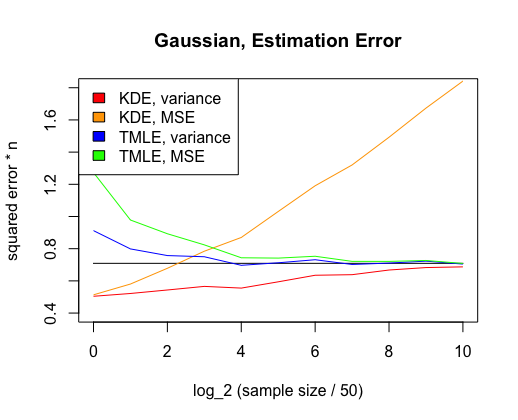}
\includegraphics[width=3 in]{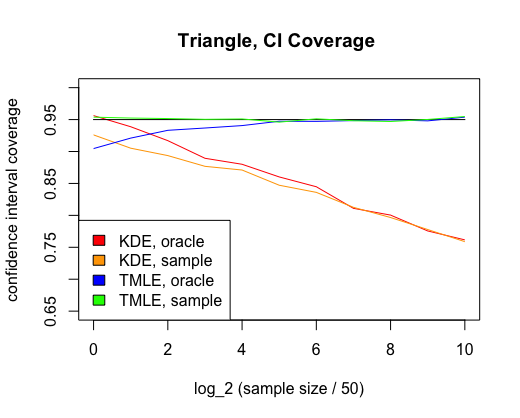}
\includegraphics[width=3 in]{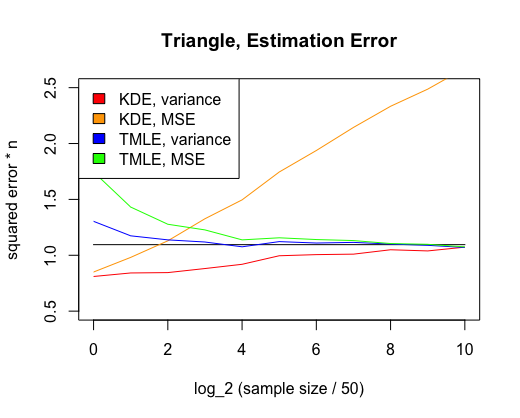}
\includegraphics[width=3 in]{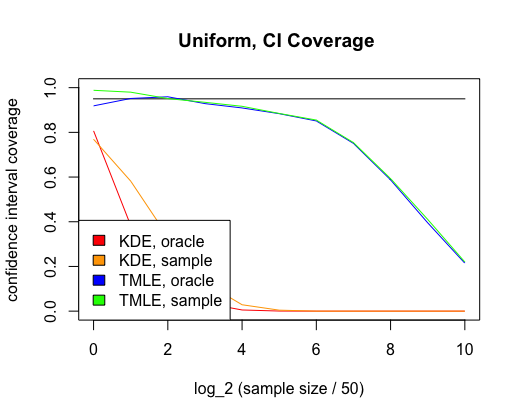}
\includegraphics[width=3 in]{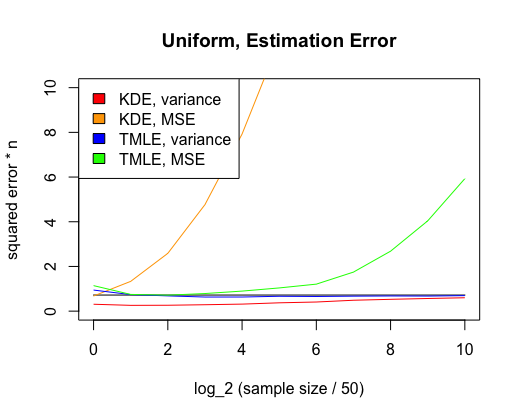}

\caption{Comparison of 95\% confidence interval coverage rate and squared error for kernel density estimator alone and TMLE. Graphs show performance for different sample sizes, where horizontal axis measures the base 2 log of sample size divided by 50. Horizontal black line shows correct coverage rate on left and variance of canonical gradient (efficient rate) on right.}

\end{figure}

\pagebreak

\subsection{Results}

We can see that for Gaussian and  right isosceles triangle, which are reasonably regular distributions, the TMLE achieves correct confidence interval coverage and is asymptotic efficient. We note that the kernel density estimator by itself does not meet these criteria. In particular, we can see the decline of the coverage rate as sample size increases, and the divergence of MSE * $n$. For the step-wise uniform distribution, which is significantly less regular, we see that the TMLE does not achieve achieve accurate coverage and is not asymptotic efficient at large sample sizes. However it's performance is even more significantly superior than that of the kernel estimator alone, compared to the other two types of simulations. This result is to be expected since the asymptotic performance of the TMLE estimator depends on the performance of the initial estimation step. The Gaussian kernel estimator is simply too poor when underlying regularity assumptions are violated. However, by looking at the mean square error, we see that it is actually in this situation that the TMLE leads to the best performance gain over the kernel alone. For smaller sample sizes we see that TMLE has higher variance and error because it fits the data. Kernel estimator alone has smaller variance and error at small sample sizes because of the degree of regularity of the distribution. We can see that the overtake point for error between TMLE and kernel alone is a larger sample size for Gaussian than for triangle, with step-wise uniform being the smallest. That tells us that the more accurate the regularity assumptions, the better the smooth kernel estimator will perform, but asymptotically the TMLE will always outperform it. 

\pagebreak

\section{Application: San Francisco Crime Incidences}

Here we demonstrate the use of L2D estimation to compare random processes in two time periods and identify the largest changes between these periods. We use publicly available data on San Francisco crime incidences (see Police Department Incident Reports). Around September 9, 2017, the San Francisco Police Department significantly increased the number of neighborhood patrol officers in several areas across the city (SFPD foot patrols). We are interested if this has had some effect on crime patterns. What types of crime are affected by this change? We will consider the geographical coordinate of crimes to be a bi-variate $X$, and our binary $A$ whether an incident occurred before or after September 9. L2D provides a way to quantify crime pattern changes across the city after the patrol officers were added. This may help identify effects resulting from policing targeted to particular locations.

Our data consists of observations of recorded crime incidents, with each observation corresponding to one incident and including the following variables: geographic coordinates of the incident, date of incident, category of crime, outcome for that incident. We compare the geographic distribution of different categories of crime in the 80 days before September 9 and the 80 days after September 9. For each category separately, $X$ is the bi-variate geographic coordinates of longitude and latitude, $A=0$ for incidents that occurred in the 80 days prior to September 9 and $A=1$ for incidents that occurred in the 80 days after September 9. Thus $p^0$ and $p^1$ are the densities of the geographic coordinates of a particular category of crime before and after September 9, respectively. We restrict our analysis to the 19 categories of crimes for which there are at least 100 incidents in both the 80 days before and the 80 days after. This is to give us a reasonable number of observations for density estimation, without using too long a stretch of time where other causes may shift the patterns of crime. For each category we estimate the L2D between the bi-variate geographic density functions of the incident generating process before versus after September 9. For each category of crime and 80 day window, we treat the recorded incident locations as $n$ iid bi-variate observations of longitude and latitude. The result is 19 L2D estimates for 19 pairs of distributions, one for each category of crime. By comparing these 19 estimates, we can identify which categories experienced the largest shift in their geographic distribution before versus after the intervention. \\

\newpage

\begin{figure}[!ht]
\begin{center}
\includegraphics[width=0.822 in]{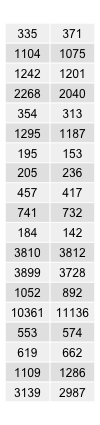}
\includegraphics[width=4.7 in]{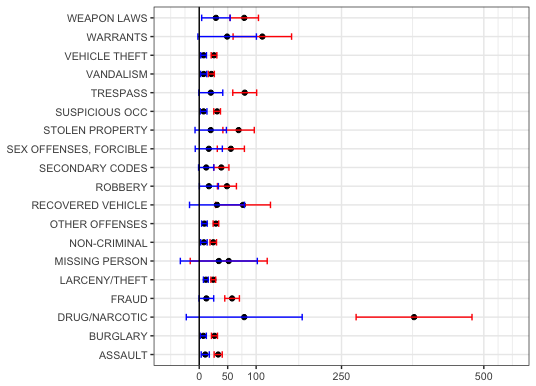}
\end{center}
\caption{\label{tab:widgets}Graph: L2D for each crime category between the density in the 80 days before the intervention and the density in the 80 days after the intervention. Two estimates are made for each category, one using kernel density estimator and the other using TMLE with kernel initial estimator. Two 95\% wald confidence intervals are provided: blue - kernel ; red - TMLE. Both use the same standard error estimate: empirical standard deviation of canonical gradient applied to kernel density estimate. Kernel density estimator use bi-variate Gaussian kernel with global bandwidth from HPI selector of Wand and Jones (1994). Table: Sample sizes (number of incidents) for each category, in the 80 days before (left column), and the 80 days after (right column) September 9.}
\end{figure}

The largest L2 distance is with drug/narcotics incidents, and it is significantly larger than the others. Further analysis shows that these types of incidents are concentrated in the downtown Civic Center area and along Mission street, a commercial avenue. These would be the places most likely to be targeted by increased police foot and bicycle patrols. Furthermore, the data shows that the vast majority of drug/narcotics incidents result in arrests, as opposed to other types of incidents which mostly have low resolution rates. This suggests
that most recorded drug/narcotics incidents correspond to interactions between police and suspects, whereas other types of recorded incidents are mostly cases of reports by third parties. It's plausible that increased foot patrols will lead to greater numbers of interactions between officers and people using narcotics in public, and this could account for the distinctive change in density for this type of incident in particular. One might expect such changes to be reflected in the number of recorded incidents across the city, with a larger percentage increase in recorded drug/narcotics incidents compared to the other categories. This need not be the case however, as the progression from summer (prior to Sep 9) to autumn (after Sep 9) could have different effects on rates of different type of crimes. Indeed we don't see a particularly large increase in the number of drug/narcotics incidents in the 80 days before to the 80 days after September 9. This illustrates an advantage of examining the probability densities and the L2 distance, which allows us to control for changes in rates affecting the entire distribution. 

Examining L2 distance can provide insights into how different random processes are affected by some change or across time. Our analysis of the SFPD data illustrates how this can be done. Changes in probability density for particular crime categories can help identify candidate categories that experienced the largest changes, and help direct further investigation. From the L2 distance, it's clear that drug/narcotics incidents stand out. Making this observation from a scatter-plot or a heat-map can be difficult. Furthermore, how the changes are distributed within categories/processes which experienced the largest overall changes can be analyzed using the difference in densities. This can be done conventionally or by the methods described by Sugiyama et al. (2013). The utility of TMLE can be seen in our analysis here as well. The coordinates of drug/narcotics incidents align closely with streets, which are lines on the map. For multivariate and/or irregularly distributed data, over-smoothing is particularly problematic for density estimation, and hence L2 distance estimation. TMLE performs targeted under-smoothing, which ameliorates this issue. We can see in this study that there can be a significant difference between kernel estimates and TMLE, so any improvement is not trivial.

\pagebreak

\section{Discussion}

In estimation we are faced with the bias variance trade off. However, the right trade off depends on what is being measured. The benefit of TMLE is that we can adjust this trade off to a particular parameter, in this case the L2D. Estimation of $\Psi(P_0)$ using a kernel smoother without TMLE update often performs poorly, especially when the underlying distribution is irregular. In particular, over-smoothing of our density estimate will bias our estimate of L2D.

When $\bold{P}_n(D^*(P_n))$ is non-zero, this term contributes to the error of our estimator. Updating $P_n$ to $P_n^*$ removes (or reduces) this error term, but at the same time we wish to avoid increasing $R^2$, which requires controlling the amount of change that we make to an accurate initial estimator. The solution is to refit, but in a manner that proportionately targets places where there is a large difference in density. The re-fitting in our update step depends across the distribution on the same $\epsilon_{MLE}$, but differs according the gradient at each location, which reflects the difference in density. This is a form of targeted under-smoothing. 

It should be noted that there is more than one solution to the problem of finding an estimate that solves its empirical influence equation. Any consistent estimator that converges faster than $n^{-1/4}$ rate will allow the TMLE to be asymptotically efficient. But at finite samples, the closer the initial estimator is to the truth, the better our performance will be. Conversely, initial estimators which fail to converge at $n^{-1/4}$ rate will not allow our TMLE to achieve asymptotic efficiency.

Finally, we will suggest some future topics of research. When using L2D to compare changes in density, it can be the case that some distributions are more concentrated than others, leading to larger squared differences when the relative changes are the same. Integrated square difference can be normalized by some appropriate function of the two densities to give a normalized distance metric that could better lend itself to comparison of random processes that have markedly different density distributions. Alternatively, instead of comparing different random processes, we could examine one random process over time. Sugiyama et al. (2013) examined the use of L2D estimation for change point detection in time series. The TMLE methodology could be used to build upon these approaches.

\pagebreak

\section{References}

\noindent
Bickel, P., \& Ritov, Y. (1988). Estimating integrated squared density derivatives: Sharp best order of convergence estimates. Sankhya: The Indian Journal of Statistics, Series A 50 381–393. \\

\noindent
Duong, T. (2007). ks: Kernel Density Estimation and Kernel Discriminant Analysis for Multivariate Data in R. Journal of Statistical Software, 21, 7.\\

\noindent
Gine, E., \& Nickl, R. (2008). A simple adaptive estimator of the integrated square of a density. Bernoulli, 14, 47-61.\\

\noindent
Kullback, S., \& Leibler, R. A. (1951). On information and sufficiency. The Annals of
Mathematical Statistics, 22, 79–86.\\

\noindent
Sugiyama, M., Kanamori, T., Suzuki, T., Christoffel du Plessis, M., Liu, S., \& Takeuchi, I. (2013). Density-Difference\\ Estimation. Neural Computation, 25, 2734-2775. \\

\noindent
van der Laan, M.J., \& Rubin, D. (2006). Targeted Maximum Likelihood Learning. U.C. Berkeley Division of Biostatistics Working Paper Series. Working Paper 213.
https://biostats.bepress.com/ucbbiostat/paper213\\

\noindent
van der Laan, M.J., \& Rose, S. (2011). Targeted Learning. Springer-Verlag New York\\

\noindent
van der Laan, M.J., \& Gruber, S. (2016). One-Step Targeted Minimum Loss-based Estimation Based on Universal Least Favorable One-Dimensional Submodels. U.C. Berkeley Division of Biostatistics Working Paper Series. Working Paper 347.
https://biostats.bepress.com/ucbbiostat/paper347\\

\noindent
van der Vaart, A.W., \& Wellner, J.A. (1996). Weak Convergence and Empirical Processes.  Springer-Verlag New York\\

\noindent
Wand, M.P., \& Jones, M.C. (1994). Multivariate Plug-in Bandwidth Selection. Computational
Statistics, 9, 97-116.\\

\noindent
Police Department Incident Reports: Historical 2003 to May 2018. Sep 12, 2018. https://data.sfgov.org/Public-Safety/Police-Department-Incident-Reports-Historical-2003/tmnf-yvry. Accessed April 1, 2019.\\

\noindent
SFPD foot patrols will quadruple in SF Mission. August 31, 2017.\\ https://missionlocal.org/2017/08/sfpd-foot-cops-will-quadruple-in-sf-mission/. Accessed April 1, 2019.\\

\pagebreak

\section{Appendix A: Derivation of canonical gradient}

$\mathbf{Theorem:}$ Consider jointly distributed $(X,A) \sim P$, $X$ multivariate continuous and $A$ Bernoulli, and parameter $\Psi(P)=\int \{p(x|A=1)-p(x|A=0)\}^2 dx $. For a model with $p^1(x) = p(x|A=1)$ and $p^0(x) = p(x|A=0)$ both non-parametric and $0<p(A=1)<1$ fixed, the canonical gradient of $\Psi$ at $P$ denoted $D^*(P)$ is: 
\begin{align*}
D^*(P)(X,A) &= \frac{I(A=1)}{p(A=1)}* 2*(p^1(X)-p^0(X) -\int (p^1(x)-p^0(x))dP^1(x)) \\
           & \quad + \frac{I(A=0)}{p(A=0)} * 2*(p^0(X)-p^1(X) -\int (p^0(x)-p^1(x))dP^0(x)) 
\end{align*}\\
$\mathbf{Corollary:}$ Define $R^2(P,P_0)$ such that: $\Psi(P)-\Psi(P_0)= (P-P_0)D^*(P) + R^2(P,P_0)$. It follows from the above that:
\begin{align*}
R^2(P,P_0)= -\int[(p_0^1(x)-p_0^0(x))-(p^1(x)-p^0(x))]^2dx
\end{align*}

\noindent
$\mathbf{Derivation:}$\\

The canonical gradient is the unique gradient of $\Psi$ that is an element of the tangent space spanned by all scores of our model.

Consider $P_\epsilon$ such that $p_\epsilon(x,a)= (1+\epsilon S(x,a))(p(x,a))$ for some score $S(x,a)$ in our model as a path in our model through $P$. The functional (or pathwise) derivative $(d/d\epsilon) \Psi(P_\epsilon)$ at $\epsilon=0$ can be expressed as a covariance under $P$ of a gradient $D$ with score $S$. The unique $D$ which can be expressed as a score within our model is the canonical gradient $D^*$\\

First we express $p(x,a)$ and $S(x,a)$ in terms of the conditional distributions and scores of $x$. Observe that: 

\[p(x,A) = p(x|A)p(A) = (p^1(x))^A(p^0(x))^{1-A}p(A).\] \\
Now define $p^1_e(x) = (1+\epsilon S^1(x))p^1(x)$ and $p^0_e(x) = (1+\epsilon S^0(x))p^0(x)$. $S^1(x)$ and $S^0(x)$ are scores for the conditional distributions with $\int S^1(x)p^1(x) dx = 0$, $\int S^0(x)p^0(x) dx = 0$. $P(A)$ is known and static in our model. Thus, $p_\epsilon(x,A) = (p^1_\epsilon(x))^A(p^0_\epsilon(x))^{1-A}p(A)$. By evaluating the derivative with respect to $\epsilon$ of $log [p_\epsilon(x,A)]$ at $\epsilon = 0$ we can express the score of $P_\epsilon$ as:

\[S(x, a) = I(A=1)S^1(x) + I(A=0)S^0(x)\]\\
Now we find $(d/d\epsilon) \Psi(P_\epsilon)$ at $\epsilon = 0$. 
 
\begin{equation*}
\begin{split}
\Psi(P_\epsilon) 
&= \int (1+\epsilon S^1(x))^2(p^1(x))^2+ (1+\epsilon S^0(x))^2(p^0(x))^2 \\
& \qquad -2(1+\epsilon S^1(x))p^1(x)(1+\epsilon S^0(x))p^0(x)  dx\\
\frac{d}{d\epsilon}\Psi(P_\epsilon(x,A))|_{\epsilon=0} 
&= 2\int S^1(x)(p^1(x))^2dx + 2\int S^0(x)(p^0(x))^2dx \\
& \quad -2\int S^1(x)p^1(x)p^0(x)dx -2\int S^0(x)p^1(x)p^0(x)dx\\
&= 2\int ((p^1(x))^2-(p^1(x))(p^0(x)))S^1(x)dx \\
& \quad +2\int ((p^0(x))^2-(p^1(x))(p^0(x)))S^0(x)dx\\
&= 2\int (p^1(x)-p^0(x))S^1(x)dP^1(x)\\
& \quad +2\int (p^0(x)-p^1(x))S^0(x)dP^0(x)
\end{split}
\end{equation*}

The above expression of the pathwise derivative can be decomposed as a covariance between a gradient and the score in our model as follows: 

\begin{equation*}
\begin{split}
&= \int 2*(p^1(x)-p^0(x))S^1(x)\frac{I(A=1)}{p(A=1)}dP(x,A)\\
& \quad +\int 2*(p^0(x)-p^1(x))S^0(x)\frac{I(A=0)}{p(A=0)}dP(x,A)\\
&= \int 2[(p^1(x)-p^0(x))S^1(x)\frac{I(A=1)}{p(A=1)} + (p^0(x)-p^1(x))S^0(x)\frac{I(A=0)}{p(A=0)}] dP(x,A)\\
&= \int [2*(p^1(x)-p^0(x))\frac{I(A=1)}{p(A=1)} + 2*(p^0(x)-p^1(x))\frac{I(A=0)}{p(A=0)}]\\
& \qquad * [S^1(x)I(A=1) + S^0(x)I(A=0)] dP(x,A)
\end{split}    
\end{equation*}

We can mean-centered both parts of the first expression in our above integrand and the value of the covariance will stay the same. We have now expressed the pathwise derivative as a covariance under $P$ of the score and a gradient. Note that this new gradient has the form of a score for the model, so it is the canonical gradient $D^*(P)(x,A)$:

\begin{equation*}
\begin{split}
D^*(P)(X,A) &= \frac{I(A=1)}{p(A=1)}* 2*\{p^1(X)-p^0(X) -\int (p^1(x)-p^0(x))dP^1(x)\} \\
           & \quad + \frac{I(A=0)}{p(A=0)} * 2*\{p^0(X)-p^1(X) -\int (p^0(x)-p^1(x))dP^0(x)\} 
\end{split}
\end{equation*}

\pagebreak

Given the canonical gradient, it is of interest to understand the error in using it as a first order approximation for the difference between the parameters of two members of our model, $P$ and $P_0$. This error is a second order term which we will refer to as $R^2(P,P_0)$. It follows from the statement of the corollary that: 

\begin{align*}
R^2(P, P_0) = \Psi(P)-\Psi(P_0) + P_0D^*(P)\\
\end{align*}

\noindent
First we evaluate $P_0D^*(P)$

\begin{align*}
P_0D^*(P) &= \int 2(p^1(x)-p^0(x) -\int (p^1(x)-p^0(x))p^1(x)dx)p^1_0(x) dx \\
        & \quad + \int 2(p^0(x)-p^1(x) -\int (p^0(x)-p^1(x))p^0(x)dx)p^0_0(x) dx\\
        & = \int 2(p^1(x)-p^0(x))p^1_0(x)dx + \int 2(p^0(x)-p^1(x))p^0_0(x)dx\\
        & \quad - \int 2(p^1(x)-p^0(x))p^1(x)dx - \int 2(p^0(x)-p^1(x))p^0(x)dx\\
        & = 2 \int (p^1-p^0)[(p^1_0-p^0_0)-(p^1-p^0)]dx\\
\end{align*}        

\noindent
It follows that:

\begin{align*}        
R^2(P, P_0) &= \int (p^1(x)-p^0(x))^2 - (p^1_0(x)-p^0_0(x))^2 dx\\
& \quad + 2 \int (p^1(x)-p^0(x))[(p^1_0(x)-p^0_0(x))-(p^1(x)-p^0(x))]dx \\
& = - \int [(p^1_0(x)-p^0_0(x))-(p^1(x)-p^0(x))]^2dx \\
\end{align*}

\pagebreak

\section{Appendix B: Asymptotic Efficiency of TMLE}

$\mathbf{Theorem:}$ Consider a TMLE $P_n^*$ such that $\bold{P}_nD^*(P_n^*)= o_P(n^{-1/2})$. Assume $D^*(P_n^*)$ is an element of a class of multivariate real valued Cadlag functions on a cube $[0,\tau]$ with sectional variation norm bounded by a universal M. In addition, assume $P_0\{D^*(P_n^*)-D^*(P_0)\}^2 = o_P(1)$ and $R^2(P_n^*,P_0)=o_P(n^{-1/2})$. Then 
$\Psi(P_n^*)-\Psi(P_0)=\bold{P}_nD^*(P_0)+o_P(n^{-1/2})$, i.e. $\Psi(P_n^*)$ is an asymptotically efficient estimator of $\Psi(P_0)$. \\

\noindent
$\mathbf{Derivation:}$\\
\noindent

Note that we can express:

\begin{equation*}
\begin{split}
\Psi(P_n^*) - \Psi(P_0)
&=-P_0D^*(P_n^*) + R^2(P_n^*,P_0)\\ 
&= (\bold{P}_n-P_0)D^*(P_0) - \bold{P}_n D^*(P_n^*)\\
& \quad + (\bold{P}_n - P_0)\{D^*(P_n^*) - D^*(P_0)\} + R^2(P_n^*,P_0)
\end{split}
\end{equation*}

$\bold{P}_n$ is the empirical measure and $(\bold{P}_n -P_0)$ is an empirical process. It follows that $(\bold{P}_n - P_0)\{D^*(P_n^*) - D^*(P_0)\}$ is $o_P(n^{-1/2})$ as a result of the asymptotic equicontinuity of an empirical process indexed by a Donsker class (van der Vaart and Wellner), where $D^*(P_n^*)$ is an element of the Donsker class of real valued Cadlag functions with bounded sectional variation norm.

Consistent density estimators will satisfy $D^*(P_n^*)$ convergent in L2(P) and any defined density is Cadlag, with $D^*(P_n^*)$ Cadlag as well. Bounded sectional variation norm requires that the density estimator is reasonably smooth. This depends on the kernel shape and bandwidth. We consider the HPI global bandwidth and other variable global bandwidth of order $n^c$. Simulations were conducted for a Gaussian kernel (not shown) and illustrated that sectional variation norm is bounded for HPI and $c= -0.3$ or greater. For $c = -0.4$ or less divergence is observed. This illustrates that for most practical density estimators the aforementioned conditions will be met. 

For $R^2(P_n^*,P_0)=o_P(n^{-1/2})$ it is sufficient that $P_n^*$ converges in L2(P) at a rate faster than $n^{-1/4}$. This depends on the choice of initial density estimator and the underlying distribution, but can be generally met by consistent density estimators.

Finally, our TMLE is a $P_n^*$ such that $\bold{P}_nD^*(P_n^*) = o_P(n^{-1/2}) $ for the one step estimator and 0 when using the universal least favorable model.

We are left with the leading term which is the empirical mean of iid random variables $D^*(P_0)(O_i)$ and is thus $n^{1/2}$ consistent and converges to a normal mean-zero limit distribution.  We have thus shown that $\Psi (P^*_n) - \Psi (P_0) $ is asymptotically equivalent to an empirical mean of the canonical gradient of our parameter, and thus it is asymptotically linear with influence curve the canonical gradient. $\Psi (P^*_n)$ is thus an asymptotically efficient estimator of $\Psi (P_0) $.

\end{document}